# Tiered Assessment in Upper-Level Undergraduate Physics

Timo A. Nieminen, Serene H.-J. Choi, and Anton Rayner
School of Mathematics and Physics, The University of Queensland,
Brisbane QLD 4072, Australia

*Abstract*

*Tiered assessment is a differentiated assessment strategy where students can choose to attempt advanced assessment tasks. We discuss the use of tiered assessment in second and third year electromagnetics courses.*

## I. Introduction

Tiered assessment is a form of differentiated assessment [1] where students can attempt a series of assessment tasks, usually of increasing difficulty. All students attempt the lowest tier of the series, but the higher tiers are optional. The lowest tier should allow students to demonstrate sufficient foundational knowledge—successful completion of this tier should be a passing score. The higher tiers allow the achievement of higher-level learning objectives to be demonstrated.

Tiered assessment can be a useful method for classes with a diverse range of student ability. As such, it has received much attention in recent years [1], especially in special education, including exceptional (i.e., high-performing) learners. However, this attention has largely been in K–12 education, and relatively little has been written concerning tiered assessment in higher education. In particular, very little has appeared on tiered assessment in physics in higher education.

We will discuss the use of tiered assessment in second and third year undergraduate electromagnetics courses. The students in these courses are, mostly, intending to major in physics or are considering doing so. Traditionally, physics has been regarded by the student body as a difficult major, and attracts high-performing students who can be passionate about the field and intend to progress to postgraduate study. Other students intend their BSc to be the terminus of their formal education. Thus, we can expect to see (and we do see in our courses) a diverse range of student ability among physics majors.

## II. Tiered Assessment

A key element of tiered assessment is the combination of basic and advanced assessment tasks. The advanced tasks should allow higher-performing students to be challenged at an appropriate level. A simple approach is to require all students to attempt all assessment tasks, both basic and advanced. However, this can greatly increase the workload of struggling students, as they attempt assessment tasks that they have little chance of completing successfully without spending excessive time on the task. In addition, if the advanced tasks are worth a large fraction of the total marks, students failing to complete them will obtain low marks, which can result in the perception of inadequate performance (even if the performance is sufficient to comfortably pass the course), with consequent increase in student anxiety and stress.

One approach to reduce the above problems is to reduce the number or difficulty of advanced assessment tasks, with a resulting reduction in the marks assigned to them. However, this reduces the motivation for learning by higher-performing students, and is undesirable. In addition, it can be difficult to distinguish between different levels of high performance if the relevant marks range is very narrow.

A more satisfactory approach is to make the advanced assessment tasks optional. This "challenge by choice" approach is the foundation of tiered assessment.

It might be argued that traditional exams are a form of tiered assessment. Typically, a result of 50% is considered a passing score, and therefore a large part of the exam is, in some sense, optional. Students will often, with the limited time available to them to complete the exam, attempt the simpler questions first, and then attempt more difficult questions if time permits. However, the problems noted above with workload and performance anxiety remain—while the exam itself remains the same length for all students, the bulk of the exam-related work consists of the preparation.

In addition, we can note a further difficulty with traditional exams: a pass mark of 50% means that it is highly likely that students can pass the exam (and perhaps the course, overall) while failing to demonstrate competence with a large portion of the course material. While the final grades for the course are typically meant to represent different levels of achievement, it is not clear that exam marks equal to the cut-offs for those final grades correspond to the levels of achievement the grades are intended to describe.

### A. Tiered Assessment and Grades

The 1–7 grading scheme with 4 passing grades is commonly used. A typical description of the levels of achievement represented by these passing grades is:

4. *Pass.* Demonstrates adequate understanding and application of the fundamental concepts.

5. *Credit.* Demonstrates substantial understanding of application of the fundamental concepts.

6. *Distinction.* As for 5, with frequent evidence of originality in defining and analysing issues or problems and in creating solutions.

7. *High Distinction.* As for 6, with consistent evidence of substantial originality and insight.

While there are 4 passing grades, there are two levels of qualitative achievement: understanding and application of fundamentals (4 and 5), and higher-level achievement (6 and 7). Grades of 4 and 5 are quantitatively differentiated (*adequate* and *substantial*), and similarly for 6 and 7 (*frequently* and *consistently*). This suggests a division of assessment tasks into two tiers, with adequate performance on the basic tier providing a minimum grade of 4 and a good performance in a minimum grade of 5. Provided that the performance on the basic tier is sufficient, the level of performance on the advanced tier will determine if a higher grade (6 or 7) is obtained.

### B. Tiered Assessment and Marks

If there is but a single assessment task for the entire course, it is sufficient to assign the final grade on the basis of performance on this assessment task. However, it is usual (or even required) to use multiple assessment tasks. It then becomes necessary to mark individual assessment tasks, and combine these marks in such a way that the final grade can be determined.

While the most common method of aggregating marks is the addition of individual marks to obtain a total mark, there are many possible methods [2], including detailed (and time-consuming) mapping schemes. The chosen method must achieve two different goals. Firstly, it must track achievement on the different tiers. Secondly, it must communicate progress to the student. Ideally, it should be easy to understand (by the students) and use (by teaching staff).

We have tested two main schemes. The assessment items typically consisted of a series of six fortnightly assignments during the semester, laboratory work, a project, and a final exam. The first scheme tried was to use to sets of marks, which we labeled "A" and "B" marks (for "advanced" and "basic"), with the final grades determined by the sums of these marks.

Since the fortnightly assignments were, individually, the smallest assessment items, these were assigned the smallest number of marks. Since we decided that there were two important levels of achievement on the basic tier (*adequate* and *complete*), we could assign 1 B mark for *adequate,* and 2 B marks for *complete* for each fortnightly assignment. Similarly, the optional advanced tier would earn 0–3 A marks. The number of marks for other assessment tasks were greater in proportion to the intended weighting. This scheme proved to have three main problems in practice. Firstly, students were confused by two sets of "marks". Secondly, the coarse marking scale for the assignments meant that a passing grade overall should be close to 50% of the total B marks. This meant that students could either perform inadequately on a large part of the basic tier, or deliberately omit a large part (this is the same as the 50% problem noted earlier). Thirdly, students performing adequately (but not higher) were obtaining about half of the B marks, and thus could view their mark as 50%, or 25% if including A marks. This led to student anxiety about low marks. The scheme did successfully track performance on the advanced tier, where the requirement of "frequently" for a grade of 6 could be met even if a large part of the advanced tier was not successfully completed (or attempted). In addition, the coarse marking scale meant that individual assessment items could be quickly marked.

The problems with this scheme suggested a two-component solution: basic and advanced tiers should be tracked in different ways, and the basic tier should be marked and aggregated in a way that strongly encouraged students to complete all of, or most of, the basic tier tasks. Since the tracking of advanced A marks appeared to work, we modified the marking of the basic tier to a grade of 1–5 (the first 5 grades on the 1–7 scale). Student performance on the basic tier was aggregated as a weighted average. The minimum performance on basic and advanced tiers for each final grade, as used in 2012, is shown in table I. The average on the basic tier required for a grade of 6 or 7 (if sufficient advanced tier marks are obtained) is below 5 to allow students to still obtain a 6 or 7 even if some grades below 5 are obtained for some individual assessment tasks.

TABLE I. FINAL GRADE CUT-OFFS

| Final Grade | Minimum Required Aggregated Results | |
|---|---|---|
| | *Basic tier weighted average* | *Advanced tier sum* |
| 7 | 4.8 | 32 |
| 6 | 4.7 | 16 |
| 5 | 4.6 | 0 |
| 4 | 4 | 0 |
| 3 | 3 | 0 |
| 2 | 1.5 | 0 |
| 1 | 1 | 0 |

Each assessment task is assigned a weighting, used to determine the contribution to the basic tier weighted average, and the number of advanced tier marks that can be earned. While for most assessment tasks, we maintain a fixed ratio of advanced tier marks to basic weighting, this is not necessary. Particular assessment tasks could be basic tier only (0 advanced marks available), or entirely optional (advanced tier only, with a weighting of 0%).

## III. ASSESSMENT TASKS

The overall structure of the assessment in 2012 for our third year course is shown in table II.

TABLE II. FINAL GRADE CUT-OFFS

| Assessment Task | Weighting and Marks | |
|---|---|---|
| | *Basic tier weighting* | *Advanced tier marks* |
| Fortnightly assignments (6 in total) | 50% in total (approx. 8.33% each) | 24 total (4 each) |
| Laboratory experiments (2 in total) | 25% in total (12.5% each) | 12 total (6 each) |
| Exam | 25% | 12 |
| Project | 0% | 20 |
| Oral exam | 0% | 20 |

The workload of the fortnightly assignments was a major factor in assigning a weighting of 50% for them in total. The fortnightly assignments are the major assessment task in the basic tier. They are also the first assessment task that students will complete and have marked. Therefore, it is important to use these to communicate the assessment scheme to students (in addition to the description of the assessment scheme in introductory materials and the course profile). Each fortnightly assignment is divided into the basic tier, labeled "Required", and the advanced tier, labeled "Additional". The "Required" section includes the instructions:

*The grade of 1–5 awarded for these exercises is based on the required exercises. If all of the required exercises are completed correctly, a grade of 5 will be obtained.*

The "Additional" section instructions are:

*Attempts at these exercises can earn additional marks, but will not count towards the grade of 1–5 for the exercises. Completing all of these exercises does not mean that 4 marks will be obtained—the marks depend on the quality of the answers. It is possible to earn all 4 marks without completing all of these additional exercises.*

We have aimed to have four questions in each 'Required' section, with the completion of three earning a grade of 4, and all four a grade of 5. Some typical basic tier questions for various topics are:

1. **Photonics:** Find the dispersion relation for TM modes in an asymmetric dielectric slab waveguide.
2. **Materials:** Calculate the group and phase velocities in the vicinity of a resonant absorption peak. What is the phase velocity for large frequencies?
3. **Relativity:** Show that the scalar product is invariant under an orthogonal transformation.
4. **Electrostatics:** Find the potential due to a dielectric sphere in a uniform applied field.

Some typical advanced tier questions are:

1. **Electrostatics:** Consider a charged conducting cube. How similar is the field or potential to that of a point charge? *Hint:* consider the symmetry of the cube and the consequent restrictions on possibly non-zero multipole moments.
2. **Photonics:** Consider a circular loop of optical fibre. How small can the loop be before bending losses become the major loss mechanism?
3. **Photonics:** Calculate the reflectivity of a layered dielectric structure as a function of wavelength.
4. **Waves:** The dispersion relation for deep water gravity-capillary waves is $\omega = (gk + (\sigma k^3/\rho))^{1/2}$ where $g$ is the gravitational acceleration, $\sigma$ is the surface tension, and $\rho$ is the density of the water. Does the dispersion curve have any interesting features? Estimate the maximum bit rate possible for the transmission of digital data over a distance of 1km.

The advanced tier questions are more difficult and more open-ended than the basic tier questions.

## IV. DISCUSSION

We believe that the current assessment scheme we are using works. The advanced tier gives students the opportunity to attempt a range of challenging problems if they choose; feedback from students is that some students enjoy the opportunity and learn a lot in the process. As many of the advanced problems are of a difficulty far in excess of the minimum requirements of the course, it is appropriate for them to be optional components of the assessment. In addition, the workload required by many students to complete the required basic tier of the fortnightly assignments is such that it is necessary for these advanced problems to be optional.

The current marking/grading scheme does not appear to result in excessive confusion. It is necessary to clearly explain it, and to remind students when appropriate, as it differs from assessment schemes they have encountered before.

Our judgment of the success of this assessment scheme is based on student feedback (formal, such as through course evaluations, and informal, such as complaints or praise) and perceived correlation between achievement of learning outcomes and final grades. We do not consider the current state of our system to be final, but still a work in progress that can be further improved.